\documentclass[10pt]{article}
\usepackage{graphicx}
\newcommand{\be}{\begin{equation}}  
\newcommand{\ee}{\end{equation}}  
\newcommand{\ba}{\begin{eqnarray}}  
\newcommand{\ea}{\end{eqnarray}}

\begin{document} 

\begin{center}
{\bf Dipolar-glass behaviour of an insulating film containing\\
 nanogranular Fe particles }\\
Norberto Majlis$^{*}$ and Martin J. Zuckermann$^{**}$\\
\end{center}
$^{*}$ Physics Department, McGill University, 3600 University St., Montreal, QC H3A 2T8, Canada.\\  
$^{**}$ Department of Physics, Simon Fraser University, 8888 University
Boulevard, Burnaby BC V5A 1S6 Canada.\\
E-mail:\\
martinz@sfu.ca\\
majlisn@physics.mcgill.ca\\

\begin{abstract}
We perform Metropolis Monte Carlo simulations of the behaviour of a film of insulating
material containing a distribution of magnetic nanoparticles. We assume that these particles only interact through dipolar forces and we find that their behaviour at low $T$ 
shows characteristics of a spin-glass with a freezing $T_f$ at which the linear susceptibility and the specific heat show a maximum. We obtain the spin glass order parameter as a 
function of temperature and we also calculate the time auto-correlation of the spin at the center of the system. We find that these results are consistent with the 
temperature dependence of the variance and the mean of the local field at the central spin.

\end{abstract}
\section{Introduction}

At present there is growing interest in low dimensional magnetic systems.
In particular,many studies are being conducted on granular systems, consisting of
nanometric clusters of a magnetic metal dispersed in a non-magnetic solid
matrix. Such systems have received much attention because of their potential application to ultra-high magnetic storage capacity.\cite{sun} 
When the matrix is electrically conducting (metal or semiconductor)
the composite system can exhibit collective behaviour due to RKKY interactions. In some cases these systems are ferromagnetic, with a critical $T_c$ in
the room temperature range \cite{ogale}.  Some of these films also display giant magnetoresistance \cite{allia}.\\
 When the host matrix is an insulator, or in general when the charge carrier
 concentration in the host is very low, the indirect RKKY
 exchange interactions between the magnetic particles can be neglected. If the average
 inter-particle separation is appreciably larger than the average particle
 size we can also neglect the effect of super-exchange, since the latter requires that the magnetic particles be in direct contact. In such a case, which is
 exemplified by some of the composite films mentioned above, the collective
 behaviour of the system is controlled by the
 magnetostatic (dipolar) interactions at low enough temperatures.\\
 Detailed numerical calculations of the RKKY
 and the dipolar interactions between clusters of varying sizes\cite{altbir} show
 that for realistic cases the latter can be as important or even more 
 important than the RKKY indirect exchange interactions for realistic situations and, furthermore,that the magnetostatic interactions can be well approximated by substituting each
 cluster by a point magnetic dipole located at its center.\\
Experimental work on the dynamics of several
different systems in which $Fe$ particles are dispersed shows that dipolar
interactions can control the behaviour of the dynamic susceptibility for an
adequate range of particle diameter and concentration. When the 
strength of the inter-particle interactions is increased, either by increasing the concentration of the particles or their radius, and hence their magnetic moment,  
the dynamic behaviour transitions from
an interaction-modified super-paramagnetism to a glassy-type collective
dynamics\cite{spinu,ohnuma,held}. \\
We have therefore chosen to study an assembly of point dipoles located at fixed
random positions inside a non-magnetic matrix. The only restriction on the space configurations is that the distance between the centres of each pair of particles is 
greater than their diameter. Their magnetic moments can orientate freely because we neglect the the effects of magnetic anisotropy.
 A Monte Carlo ($MC$) simulation of a planar triangular lattice of nanoparticles with random anisotropy and 
interacting through dipolar forces has shown the interplay of both energies.\cite{wagner} 

In this article we present the results of Metropolis Monte Carlo ($MMC$) simulations for a thin film
containing  a completely random but fixed spatial distribution of point particles, representing bcc Fe clusters, which only interact through magnetic dipolar forces. 
The $MMC$ algorithm itself is described in the next section and the simulation data is exhibited and discussed in the following sections. Emphasis is placed on the spin glass nature of the results.
\section{Model and Method}
We incorporate the periodic boundary conditions in the film by introducing a square lattice of cells and  we then randomly distribute
$ N=n_x^2\,n_z$ particles inside each cell. Here $n_y=n_x$. This basic cell is then
repeated indefinitely along the $(x,y)$ plane. The total dipolar energy  of the film is given by:
\begin{equation}
W=1/2\sum  _{{\bf r}_n}\sum _{i,j=1}^{N}
{\bf \mu} _i \cdot {\bf D} ({\bf R} _{ij}+{\bf r}_n) \cdot {\bf \mu}_j
\nonumber
\end{equation}
where ${\bf r} _n $ is a site in the square $(x,y)$ lattice and the $\alpha , \beta $ components of the dipolar tensor ${\bf D}$ are
$$ D^{ \alpha , \beta }({\bf  R})=-\nabla _{R_{\alpha}}\nabla _{R_{\beta }}
\frac{1}{\mid {\bf R} \mid }$$
 Here frustration, which is a necessary ingredient for glassy behaviour, is a consequence of the disorder in the positions and orientations of the dipoles, 
resulting in random sign and amplitude fluctuations of the tensorial dipolar interaction. 
We have chosen a fixed value of $n_z = 3$ , which corresponds to a width of the film equal to $3 \times f\times d$ ,where $d=10\AA$ is taken to be the particle diameter and 
$f>1$ is a factor which determines the average inter-particle distance. We
choose $f=2$.\\

Dipolar sums are calculated by
the adaptation of {\it Ewald's} summation algorithm to the quasi-two-dimensional case and the $MMC$ algorithm is used to calculate both thermodynamic and local properties of the system.
All quantities are averaged for every temperature over 100 -200 different random space configurations of the particles. The $MMC$ runs consisted of $80000$ steps for the 
warming cycle, and $40000$ steps for the actual calculation.

We place  a particle at the center of the film cell in all cases and we calculate the average and the variance of the components of the 
local field ${\bf B} _0$ and of the central spin ${\bf S} _{0}$ for this site.
In addition the following physical variables are calculated as functions of temperature for different numbers of
particles in the basic cell: the total energy per particle, the susceptibility tensor and
the specific heat. Their values are then extrapolated to obtain the
limit of an infinite film of $n_z=3$, namely the width corresponding to $3$ average distances.
We also obtain the averaged time auto-correlation function $g(t)$ of the central particle spin ${\bf S} _0$, defined as
\be
g(t)=\langle\langle {\bf S} _0(t_0)\cdot {\bf S} _0(t_0+t)\rangle\rangle _{t_0} 
 \label{eq:gedete}
\ee
where an average over the initial time, $t_0$, is performed for each statistical average.
Note that in our calculations time, $t$, is defined as the number of $MMC$ steps as the simulation progresses. This implies from the definition above that we calculate the scalar 
product of the value of central spin, ${\bf S}_0$, after $t_0$ MC steps with
the value of the central spin after a further $t$ $MMC$ steps. The time scale is not otherwise defined, 
being dependent on the spin-flip physical time, which in real systems is of the order of $10^{-13} \,sec$.
From the original Anderson definition of the order parameter $q_{EA}$ \cite{edwards} we have:
\be
q_{EA}=\lim  g(t)\mid _{t \rightarrow \infty} \label{EA}
\ee
An alternative order parameter has been defined as follows \cite{diep}:
\begin{equation}
q_{H}=\frac{1}{N}\sum _{i=1}^{N}\left(\sum_{\alpha=x,y,z}\,\vline \frac{1}{\tau}\,\,\sum
_{t^{\prime}=t_{w}}
^{t^{\prime}=t_{w}+\tau }\,\,S^i_{\alpha}(t^{\prime})\,\vline\,^2\right)^{1/2}
\nonumber
\end{equation}
We calculate both order parameters for every value of the temperature $T$.

\section{Thermodynamic properties}
\begin{figure}[h!]
\begin{center}
\includegraphics*[height=6.5cm]{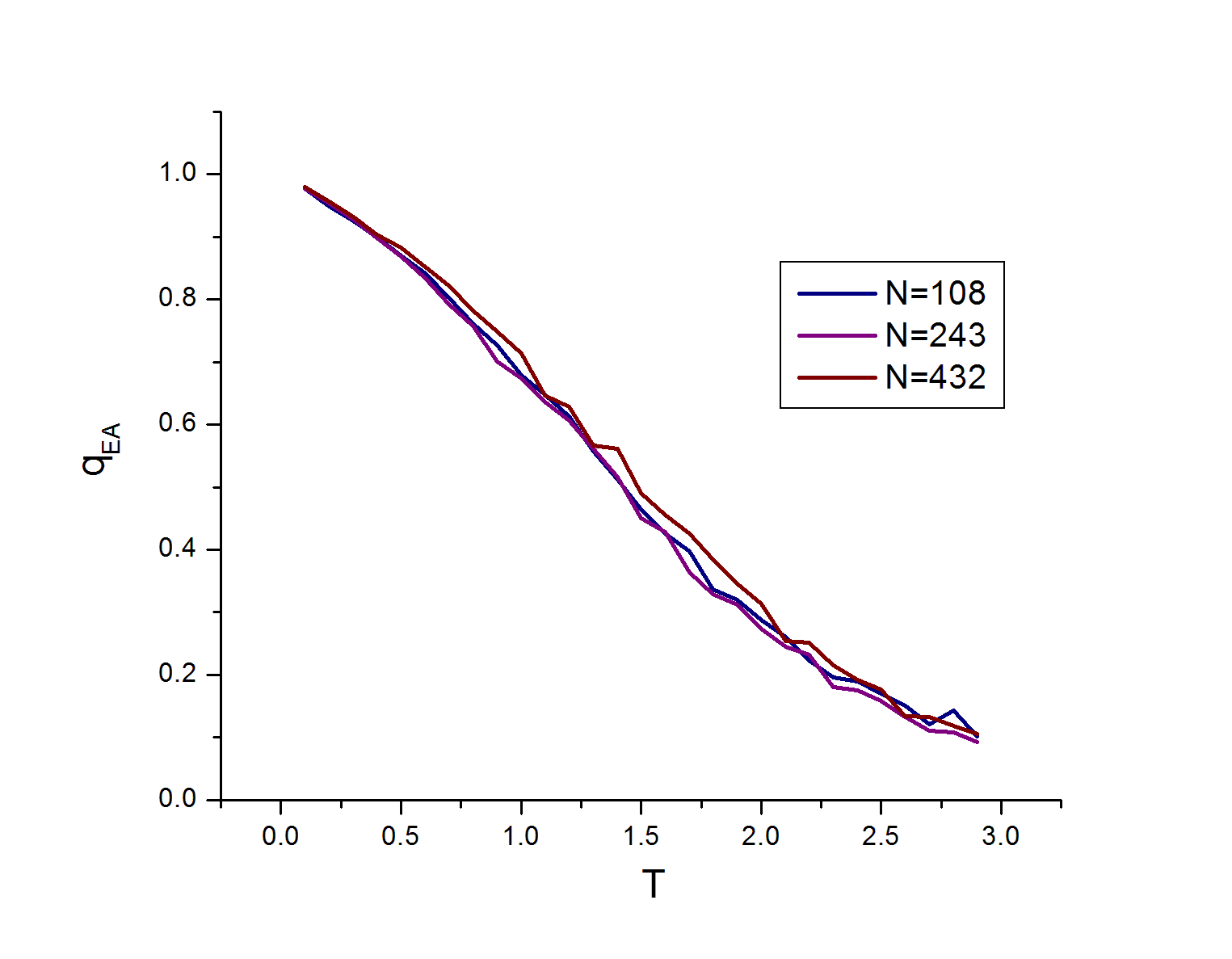}
\caption[q_{EA} order parameter]{{\it The order parameter $q_{EA}$ as a
 function of $T$ in degrees Kelvin for three different sizes.}}\label{qEAM}
\end{center}
\end{figure}

\begin{figure}[h!]
\begin{center}
\includegraphics*[height=6.5cm]{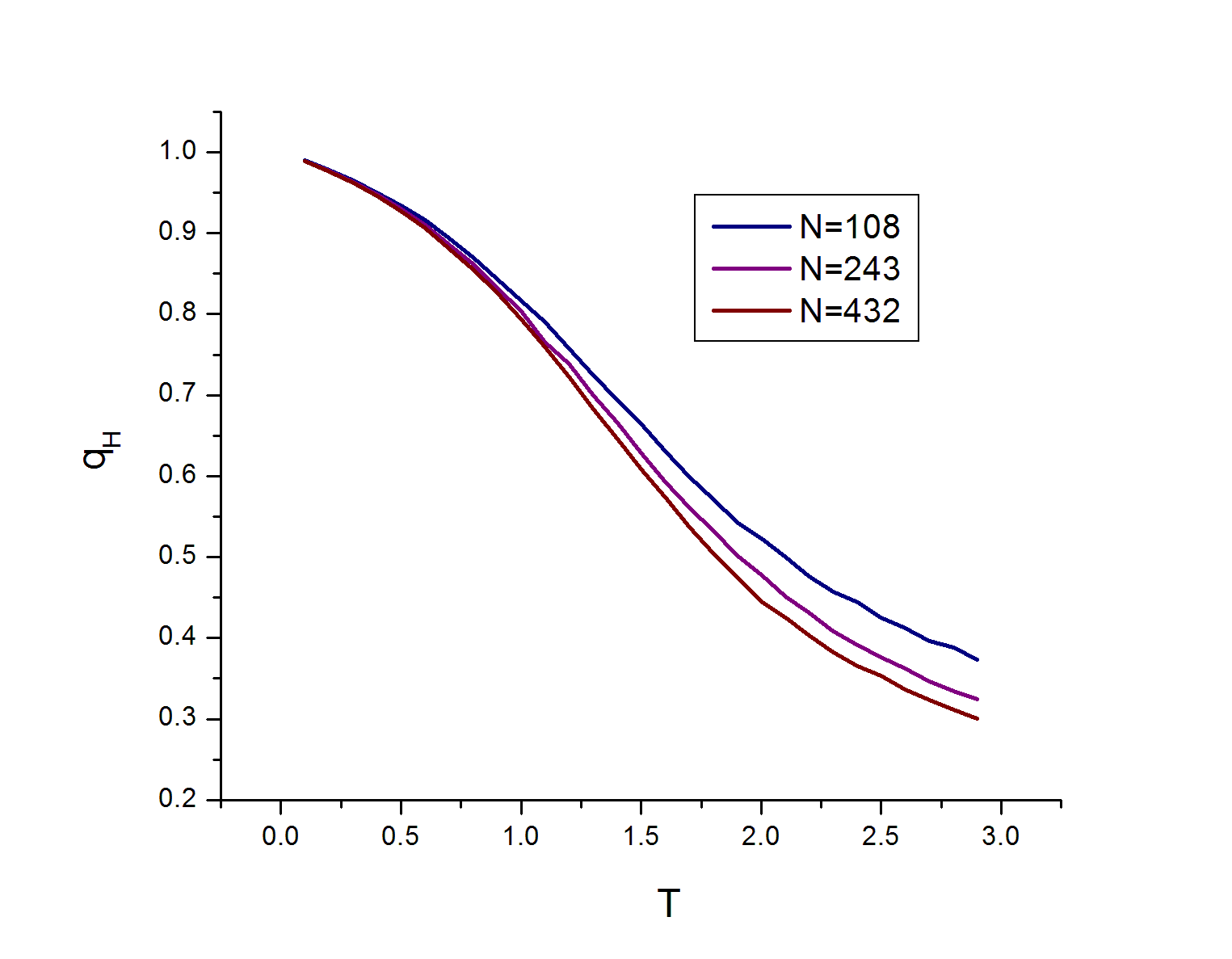}
\caption[q_{H} order parameter]{{\it The order parameter $q_H$ as a
 function of $T$ for different sizes. }}\label{qH}
\end{center}
\end{figure}

At $T=0$ order parameters defined in the previous section should equal $1$ 
while they are expected to vanish above the freezing temperature $T_{f}$. 
 However due to the system's finite size, 
both  $q_{EA}(T)$ and $q_H(T)$ have a long tail for high $T$ as can be seen in figures \ref{qEAM} and \ref{qH}.\\

One can fit the curves for $q_{EA}$  and $q_{H}$ vs. $\,T$  with
an algebraic function such that they intersect the temperature axis at an extrapolated  temperature $T^{(\alpha)}_{q}(N)$ ( $\alpha = \{ EA,H \}$) 
which is an estimate of the transition temperature from the the spin-glass state to the disordered
(super-paramagnetic) phase, for every value of $N$. We fit the low $T$ curves for both order parameters with the function:
\begin{equation}
f(T)=(1.0-T/T_q)^{\nu} \label{fit}
\end{equation}
which for low $T$ behaves as
\begin{equation}
f(T)\approx 1.0 -\nu \,T/T_q \label{fitEA}
\end{equation}

\begin{figure}[h!]
\begin{center}
\includegraphics*[height=6.5cm]{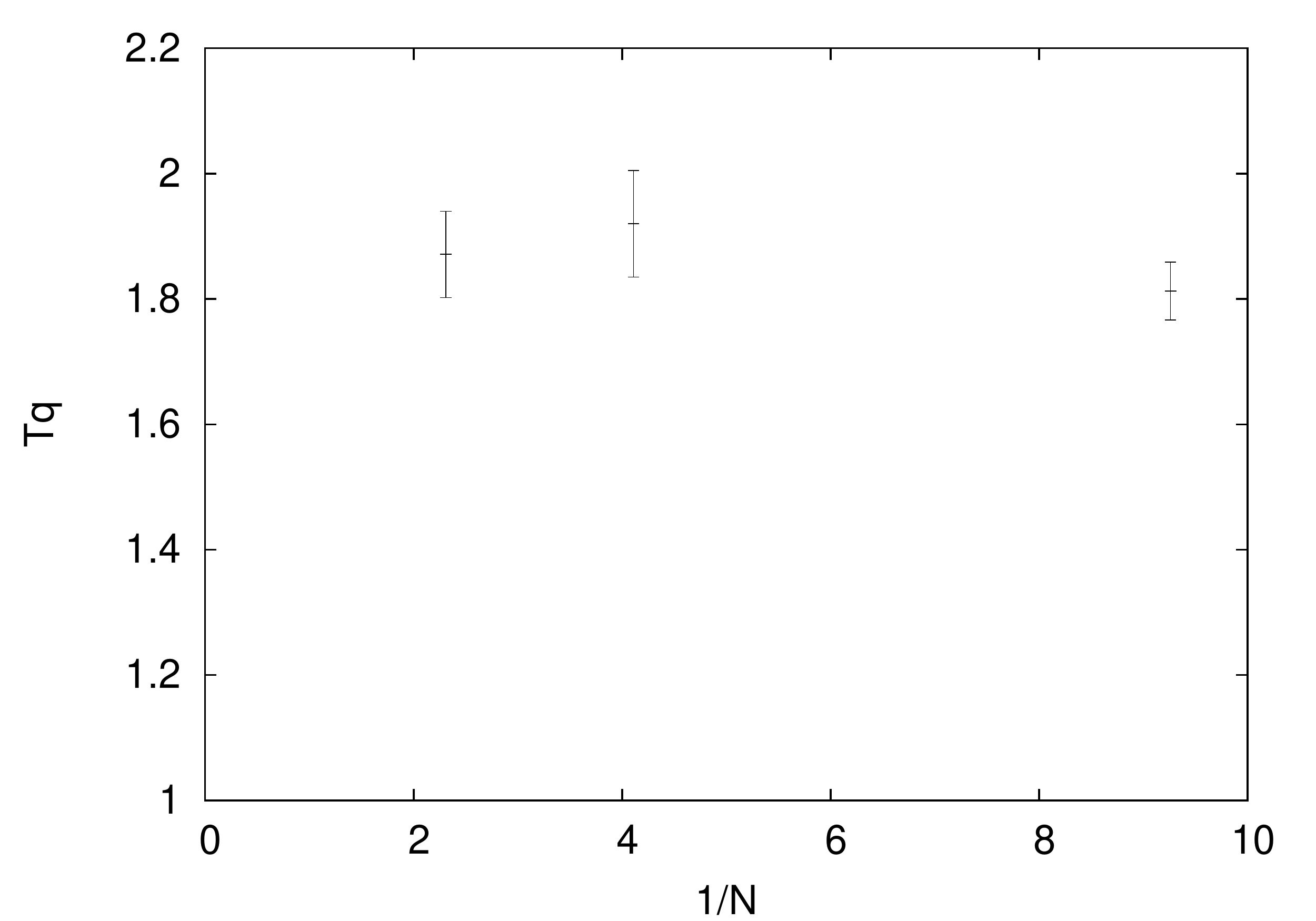}
\caption[Extrapolated T_q]{{\it Extrapolated $T_q$ in Kelvin as a function of
 $\frac{10^3}{N} $.}}\label{Tq}
\end{center}
\end{figure}

The freezing temperatures thus obtained for both order parameters are in good agreement with each other, indicating that both definitions are consistent. However, $\it q_{EA}$ is 
amenable to a better extrapolation, since it is less sensitive to the finiteness of the sample used in the simulation. At any rate, one has to  cutoff the data at a reasonable value of $T\approx 1.4-1.6 K$,
 since as mentioned above the order parameters for a finite system do not vanish at any finite $T$. This arbitrary
 procedure introduces an uncertainty in $T_q$ of the order of $0.1-0.2 K$. \\ 
In Fig. \ref{Tq}  we show extrapolated values of $T_q$ for three different values of $N$. From these data we estimate:
 $$\lim_{N\rightarrow \infty}T_q \approx 1.79 K \pm .07 $$
 $$\lim_{N\rightarrow \infty}\nu \approx 0.42 \pm .07$$
so that at low T we get
\be
q_{EA}\approx 1.0 -0.42 T/T_q  \label{fitlow}
\ee 
It is noteworthy that the mean field solution of the $EA$ model of a spin glass, as mentioned by Mydosh \cite{Myd} yields at low $T$ :
\begin{equation}
q_{EA}(T)=1.0 - 0.4066 T/T_f \label{eq:fitEA}
\end{equation}
 in close agreement with our result in Eq. \ref{fitlow}.

 We also obtain an estimate of $T_f $
 by calculating, following {\it Binder},\cite{binder}, the kurtosis of the distribution of the  total magnetization as a function of $T$, defined as 
 \be
 cumul(T) = 1.0 - \langle \vline \vec{ M}\vline ^4\rangle /3\langle\vline \vec{ M}\vline ^2 \rangle ^2  \label{bindeq}
\ee

The temperature at which the plots of this quantity vs. $T $ intersect for different values of $N$ should give an estimate of the freezing temperature for infinite size. 
It should be noted that the kurtosis is zero for a Gaussian distribution. \\

\begin{figure}[h!]
\begin{center}
\includegraphics*[height=6.5cm]{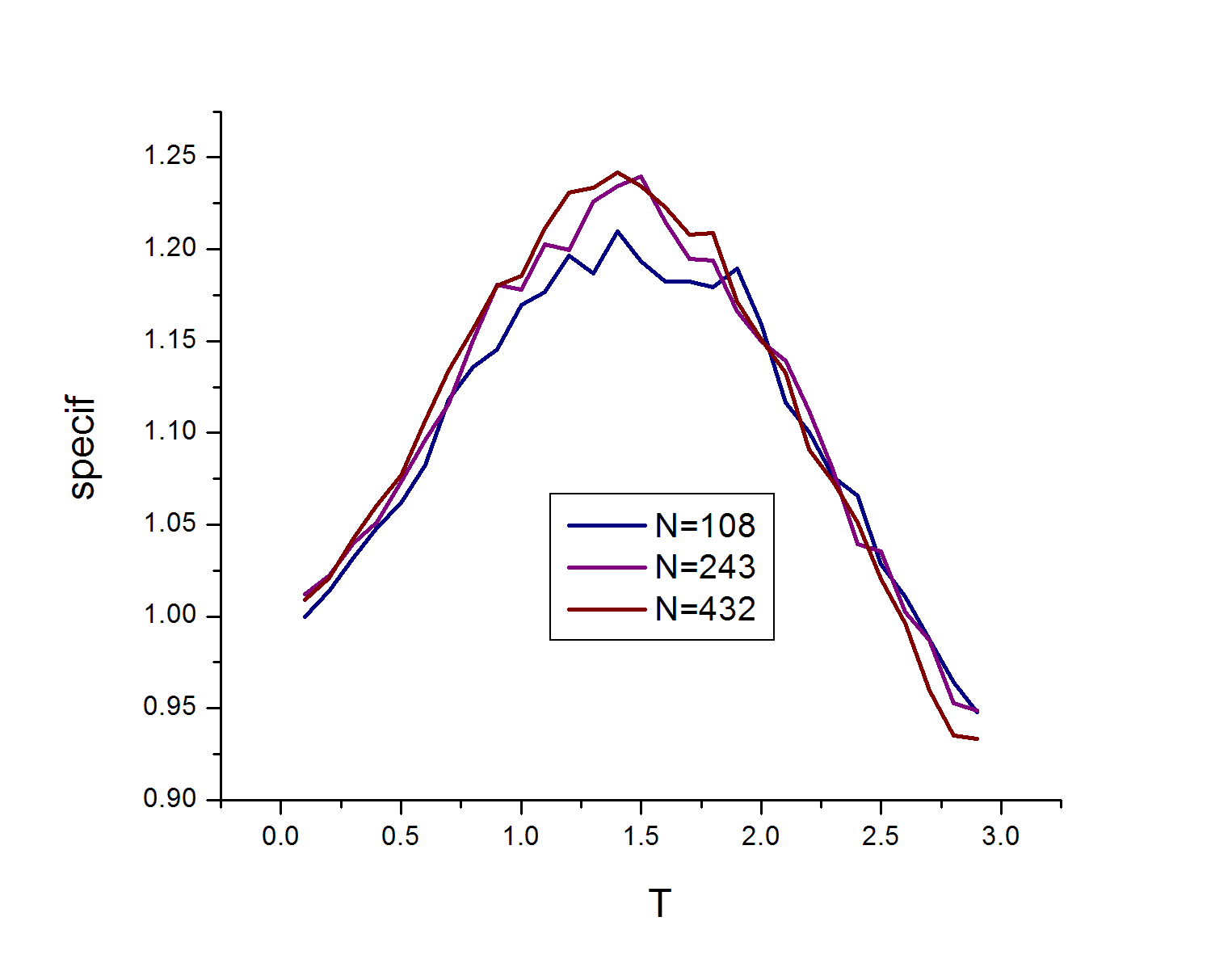}
\caption[Specific Heat]{{\it The Specific Heat (variance of the energy) as a
 function of $T$ in Kelvin.}}\label{specif}
\end{center}
\end{figure}

\begin{figure}[h!]
\begin{center}
\includegraphics*[height=6.5cm]{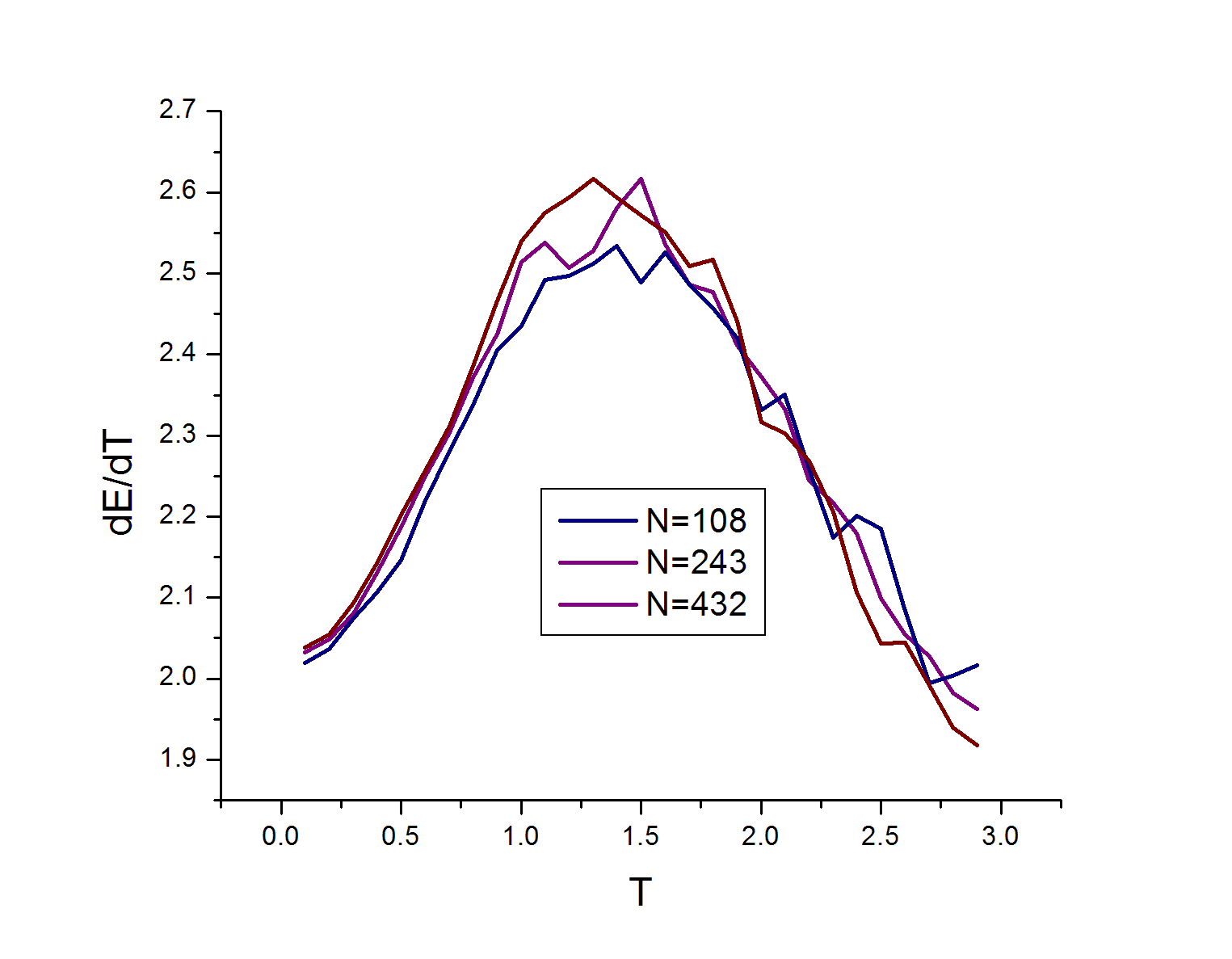}
\caption[Specific Heat]{{\it The Specific Heat (Temperature derivative of the energy) as a
 function of $T$}}\label{ederiv}
\end{center}
\end{figure}

\begin{figure}[h!]
\begin{center}
\includegraphics*[height=6.5cm]{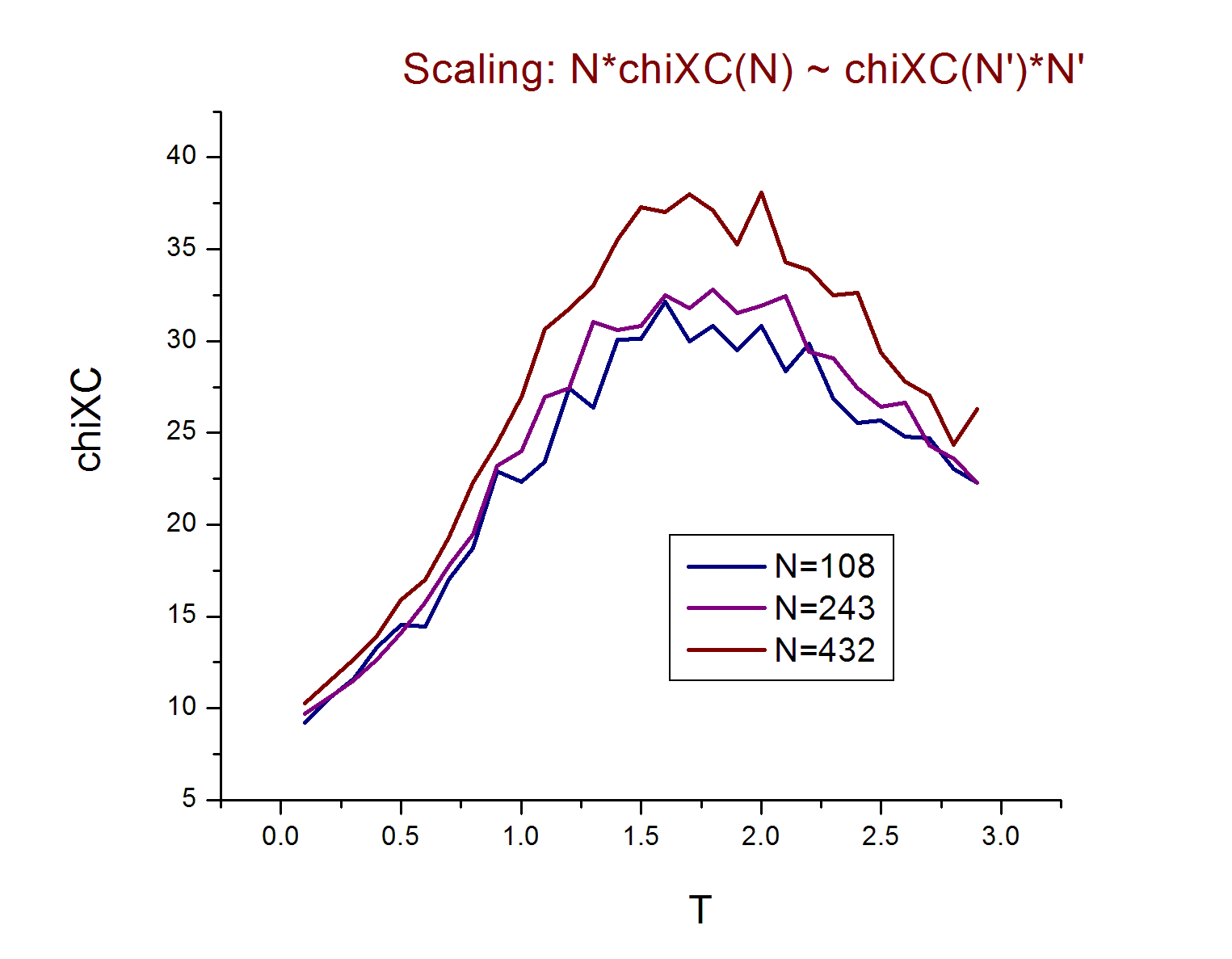}
\caption[Susceptibility]{{\it The Static Longitudinal Susceptibility as a
 function of $T$}}\label{sus}
\end{center}
\end{figure}

The specific heat is shown in Fig. \ref{specif} as the variance and in Fig. \ref{ederiv} as the numerical derivative, of the energy. 
Just as for the  parallel static susceptibility curve in Fig. \ref{sus} they
show a maximum at a temperature $T_m(N)$ slightly lower than $T_q(N)$. These  results are compatible with spin-glass behaviour.\cite{raju}\\

\begin{figure}[h!]
\begin{center}
\includegraphics*[height=6.5cm]{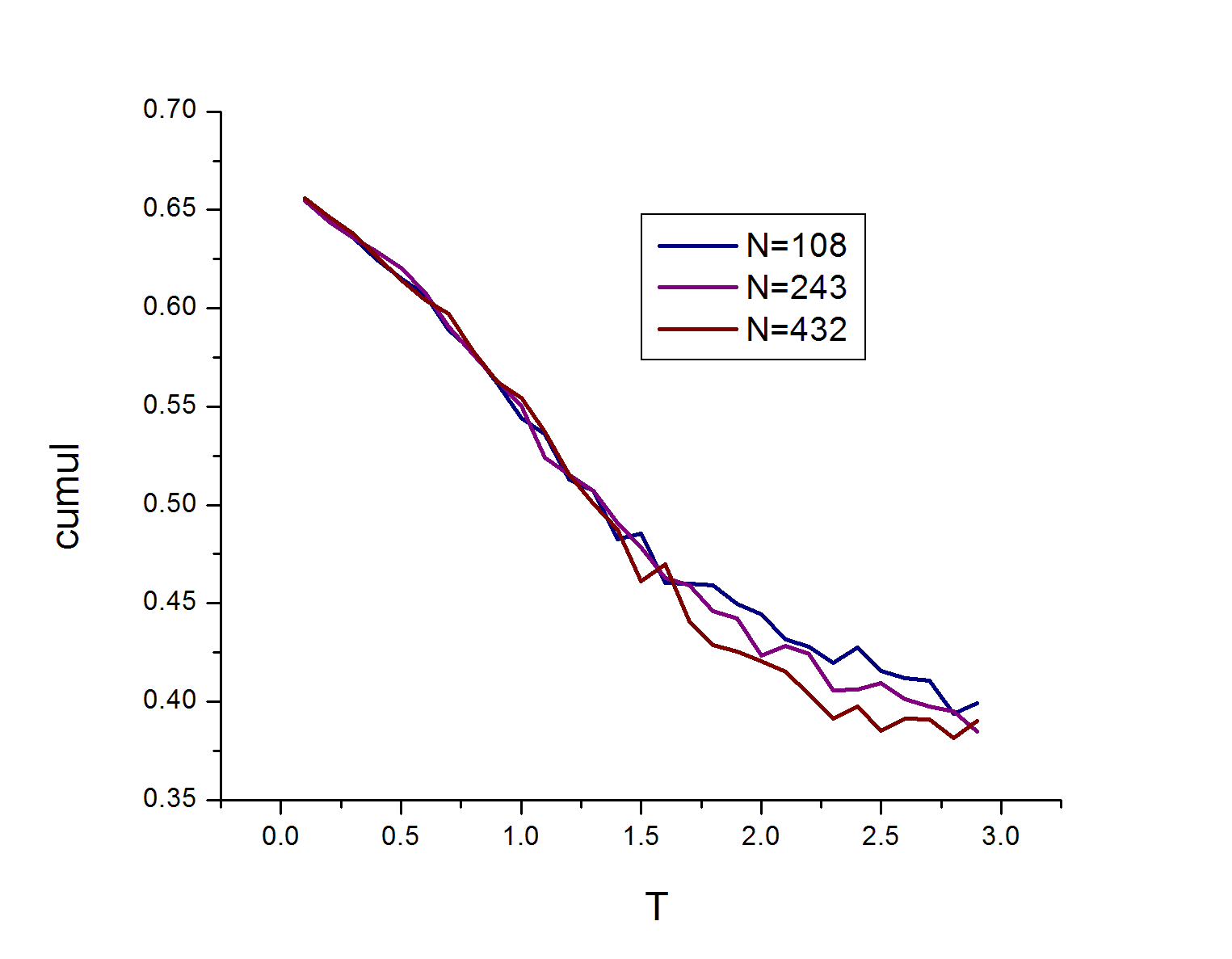}
\caption[Binder cumulant]{{\it Binder cumulant expression vs. $T$ for different sizes.}}\label{cumul}
\end{center}
\end{figure}

The graphs of the  Binder cumulant combination, defined in Eq.\ref{bindeq}, as a function of $T$ 
for three different sizes intersect approximately at a temperature $T_c$, which is considered to be an estimate of $T_f$.
From Fig.  \ref{cumul} we obtain $ T_c \approx 1.5\, K$. 

\begin{figure}[h!]
\begin{center}
\includegraphics*[height=6.5cm]{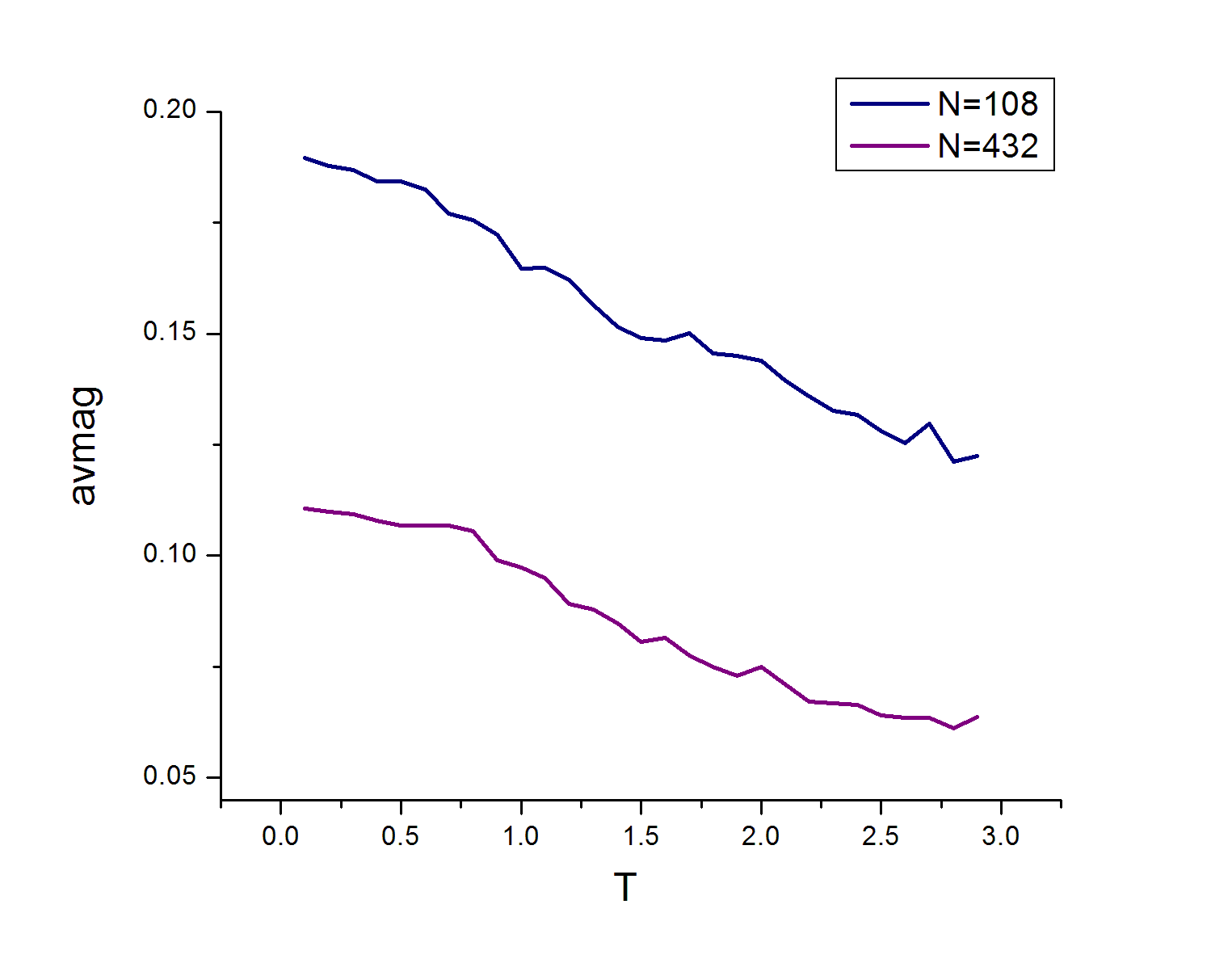}
\caption[Magnetization]{{\it Magnetization per particle vs. $T$ for different sizes.}}\label{avmag}
\end{center}
\end{figure}

We verify (Fig. \ref{avmag}) that the average modulus of the magnetization scales as $N^{-1/2}$ as correspods to a completely random dipole distribution.

\begin{figure}[h!]
\begin{center}
\includegraphics*[height=6.5cm]{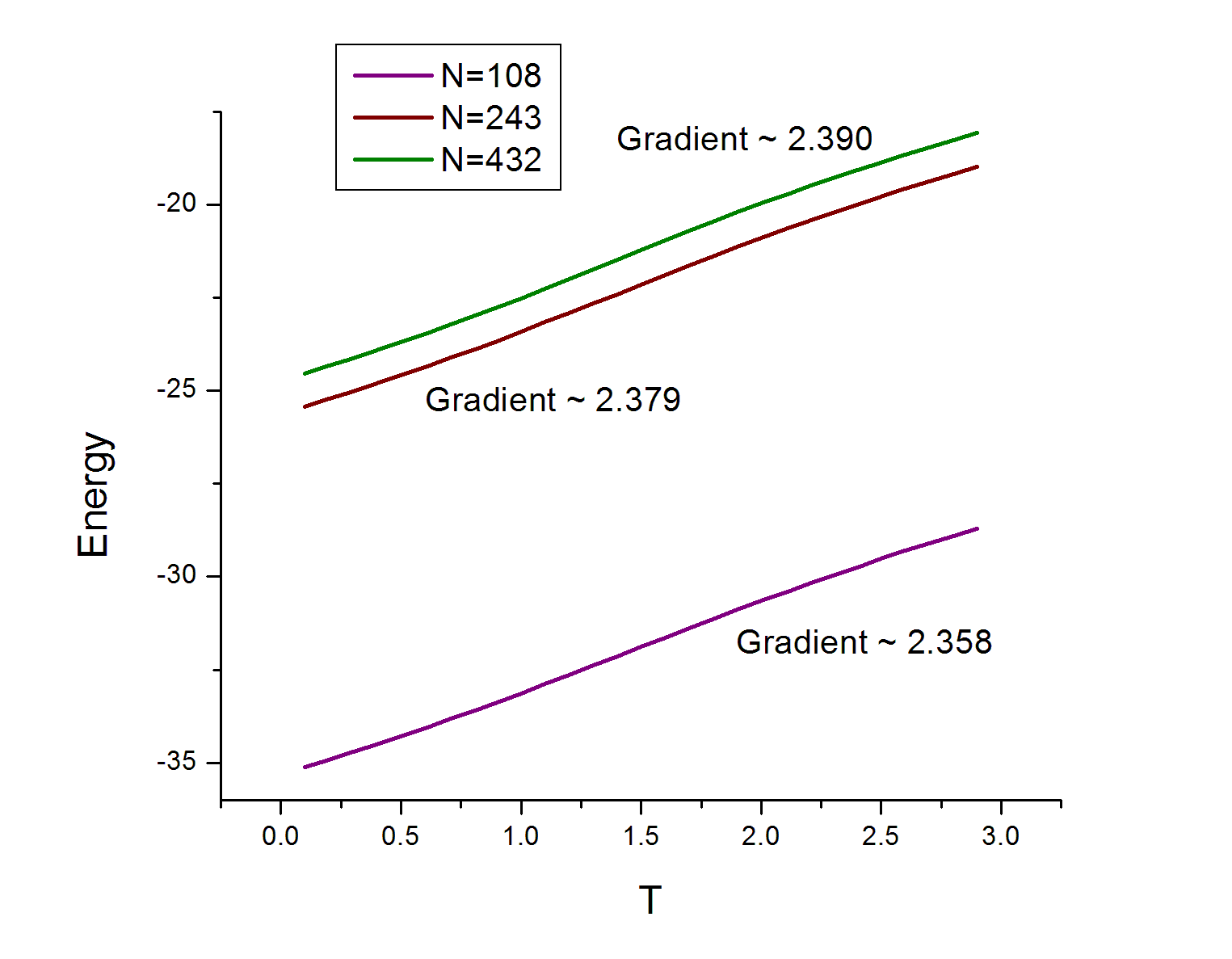}
\caption[Energy]{{\it Dipolar energy per particle vs. $T$ for different values of $N$.}}\label{energy}
\end{center}
\end{figure}

The energy per particle is shown in Fig. \ref{energy} for three different values of $N$. We verify that it converges as $N$ increases.

\section{Local properties}

\begin{figure}[h!]
\begin{center}
\includegraphics*[height=6.5cm]{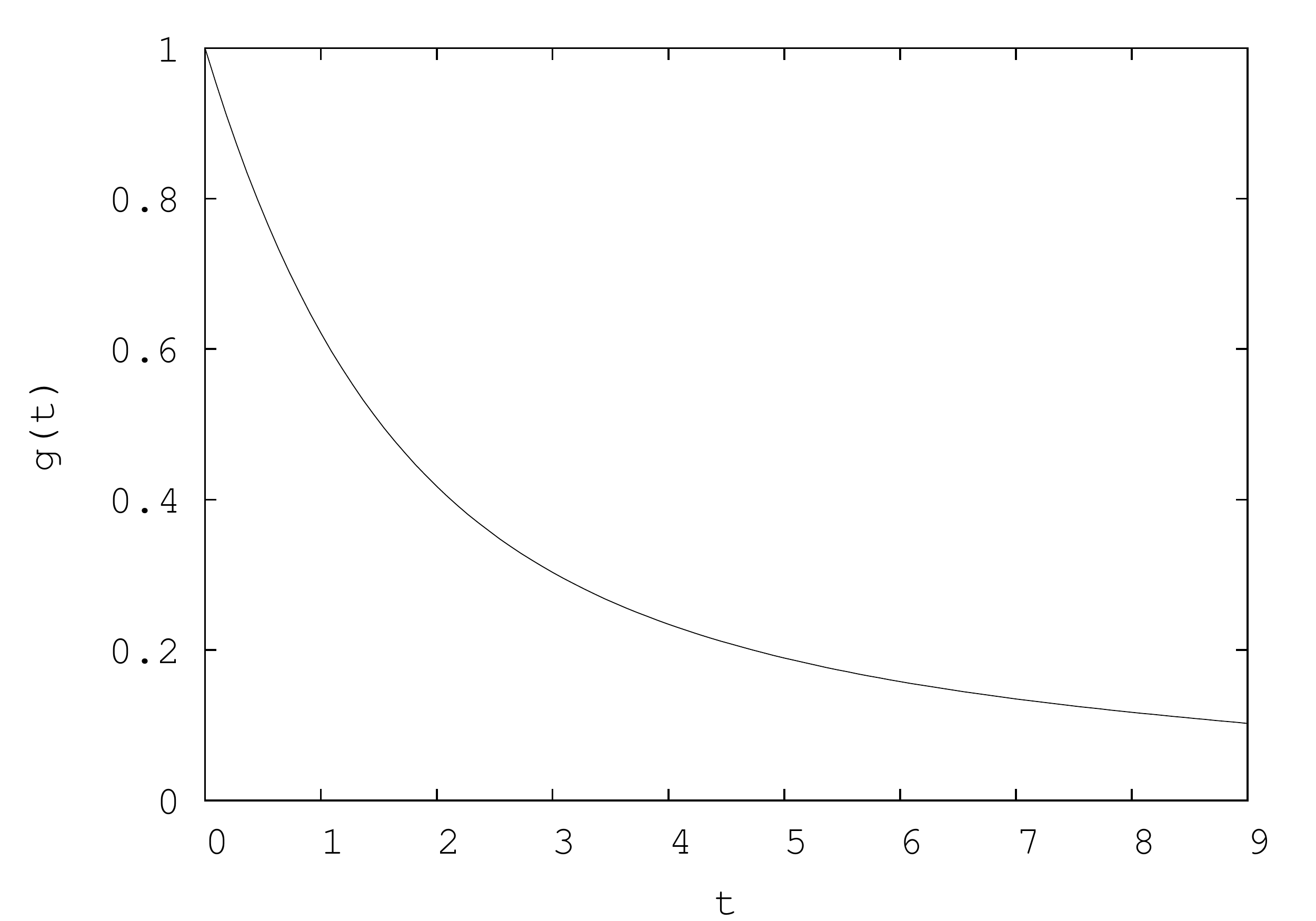}
\caption [Time Self-correlation.]
{{\it  Time Self-correlation of central spin $g(t)$.}}\label{corr}
\end{center}
\end{figure}

Besides  the thermodynamic properties we have studied several local characteristics of the system.\\
The time auto-correlation function $g(t)$ mentioned above was obtained for the central spin. A plot of  $g(t)$  for $N=108$,  $T=2.9 K$ and 160 configurations
is shown in Fig. \ref{corr}. By definition $g(0)= 1$ at all temperatures.

\begin{figure}[h!]
\begin{center}
\includegraphics*[height=6.5cm]{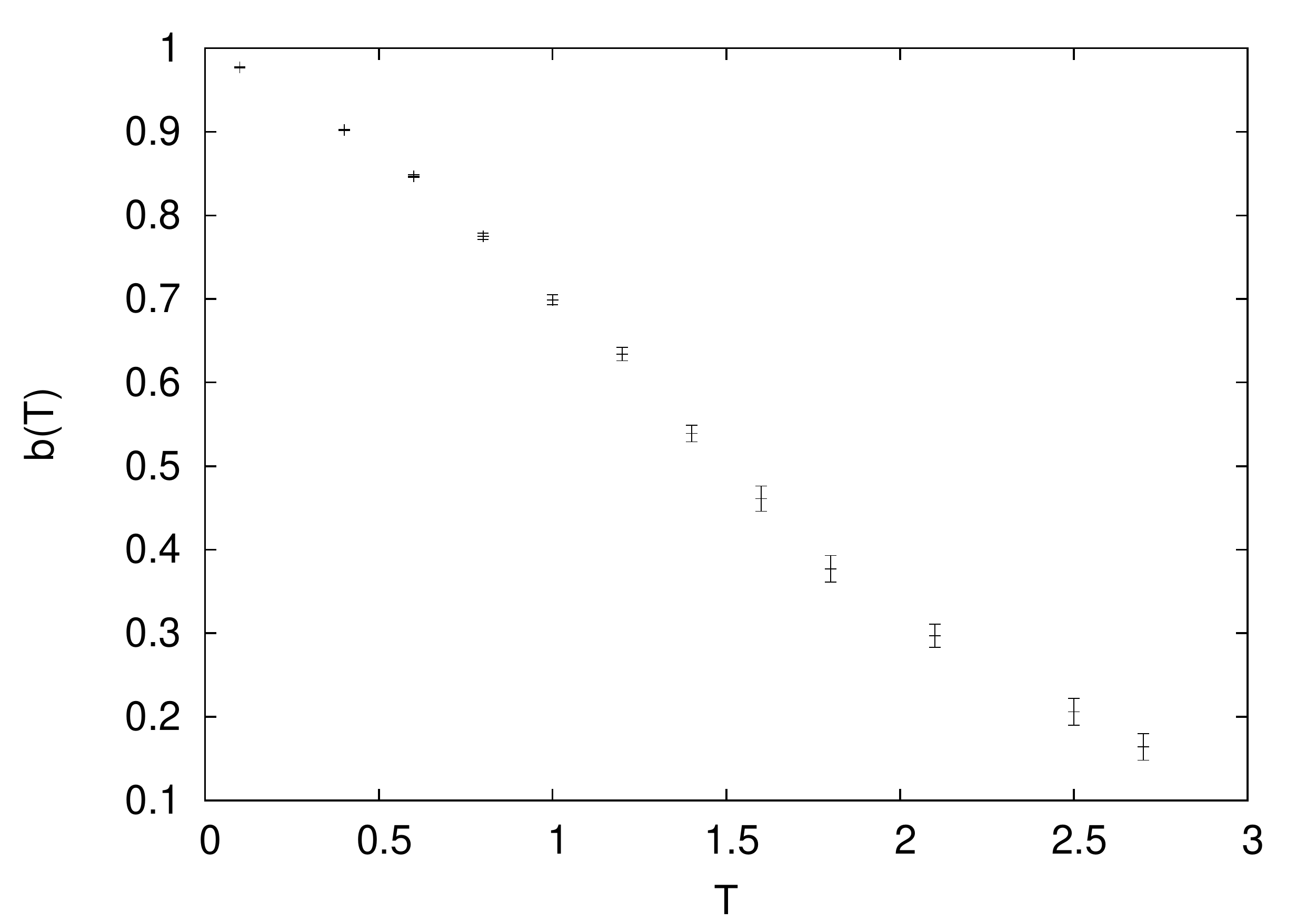}
\caption [The function $b(T)$.]
{{\it The function $b(T)$ for N=108.}}\label{bT}
\end{center}
\end{figure}

\begin{figure}[h!]
\begin{center}
\includegraphics*[height=6.5cm]{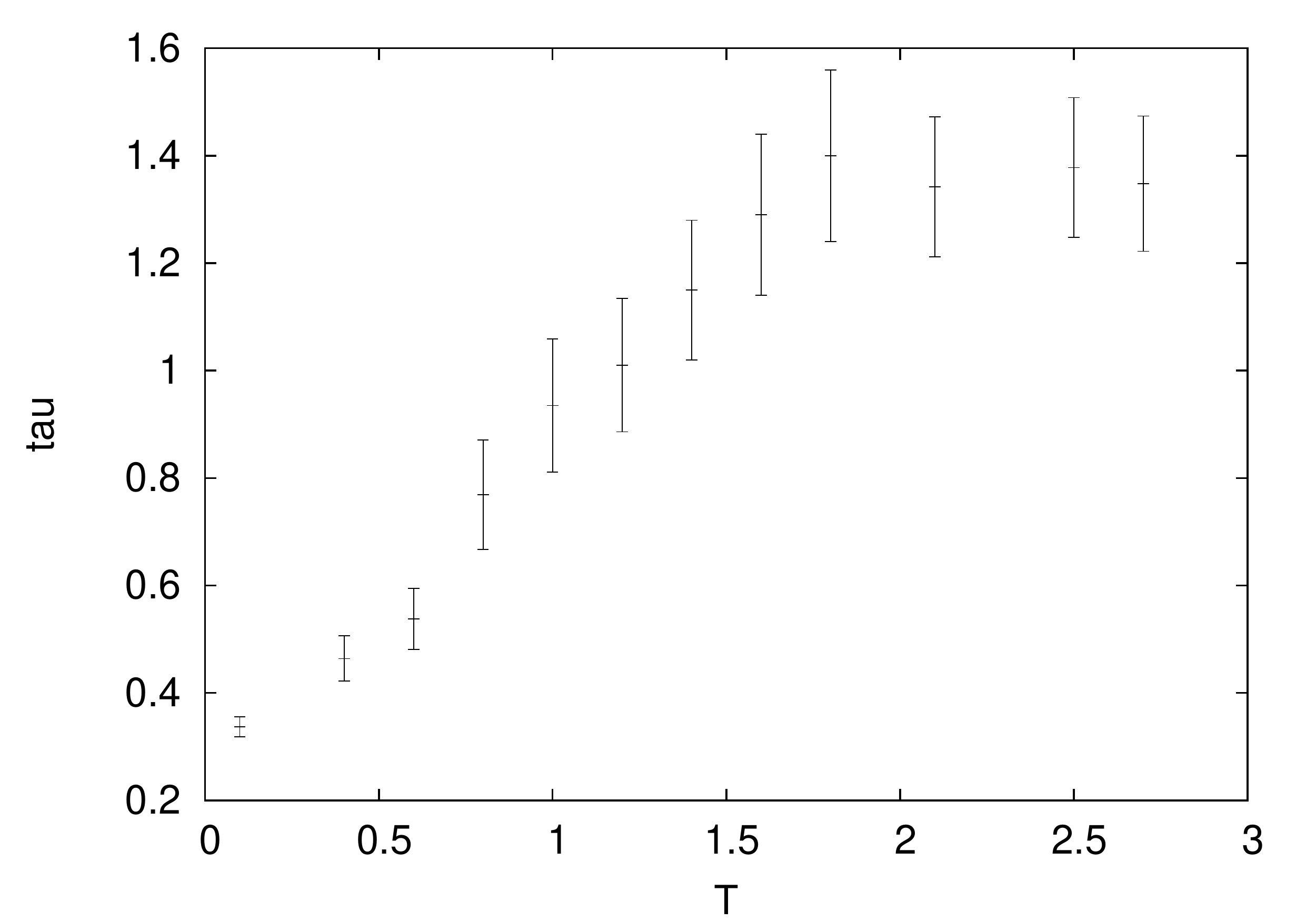}
\caption [Decay Time of Self-correlation, $\tau (T)$.]
{{\it Transient decay time $\tau (T)$ of central spin self-correlation.}}\label{tauM}
\end{center}
\end{figure}

The best fit to $g(t)$ was found to be 
\be
g(t)= b(T)\,+ (1.0-b(T))\exp {(-t/{\tau (T)})}  \label{fit:gedete}
\ee
where, as indicated above, both parameters are functions of $T$.
Our results for $g(t)$ at different temperatures allow us to obtain the functions $b(T)$ and $\tau (T)$, shown in Figs. \ref{tauM} and \ref{bT} 
with the standard statistical error (SSE) bars of the fit.

From Eq. \ref{EA} we must have $$b(T)=q_{EA}(T)$$
which we verify by comparing with Fig. \ref{qEAM}. This is a consistency requirement on the fit to $g(t)$.
The characteristic decay time $\tau$ of the transient term in Eq. \ref{fit:gedete} is shown as a function of $T$ in  Fig.\ref{tauM}.

We see that as $T\, \rightarrow 0$  $g(t)$ tends to $1$ while the decay time of the transient vanishes. As $T$ increases above $T_f$ the transient gets longer and the limiting amplitude decreases, 
and should eventually vanish. 
One again finds an abrupt change in the derivative of the curve for $\tau(T)$
 which is the signal of the freezing temperature.\\  
 As a necessary step in the calculation of the energy and the specific heat we need to obtain the local field on every particle. 
The statistical distribution of the values of all three components of the field can be obtained as the {\it MMC} run proceeds, and we 
obtained their average and variance from this distribution.

\begin{figure}[h!]
\begin{center}
\includegraphics*[height=6.5cm]{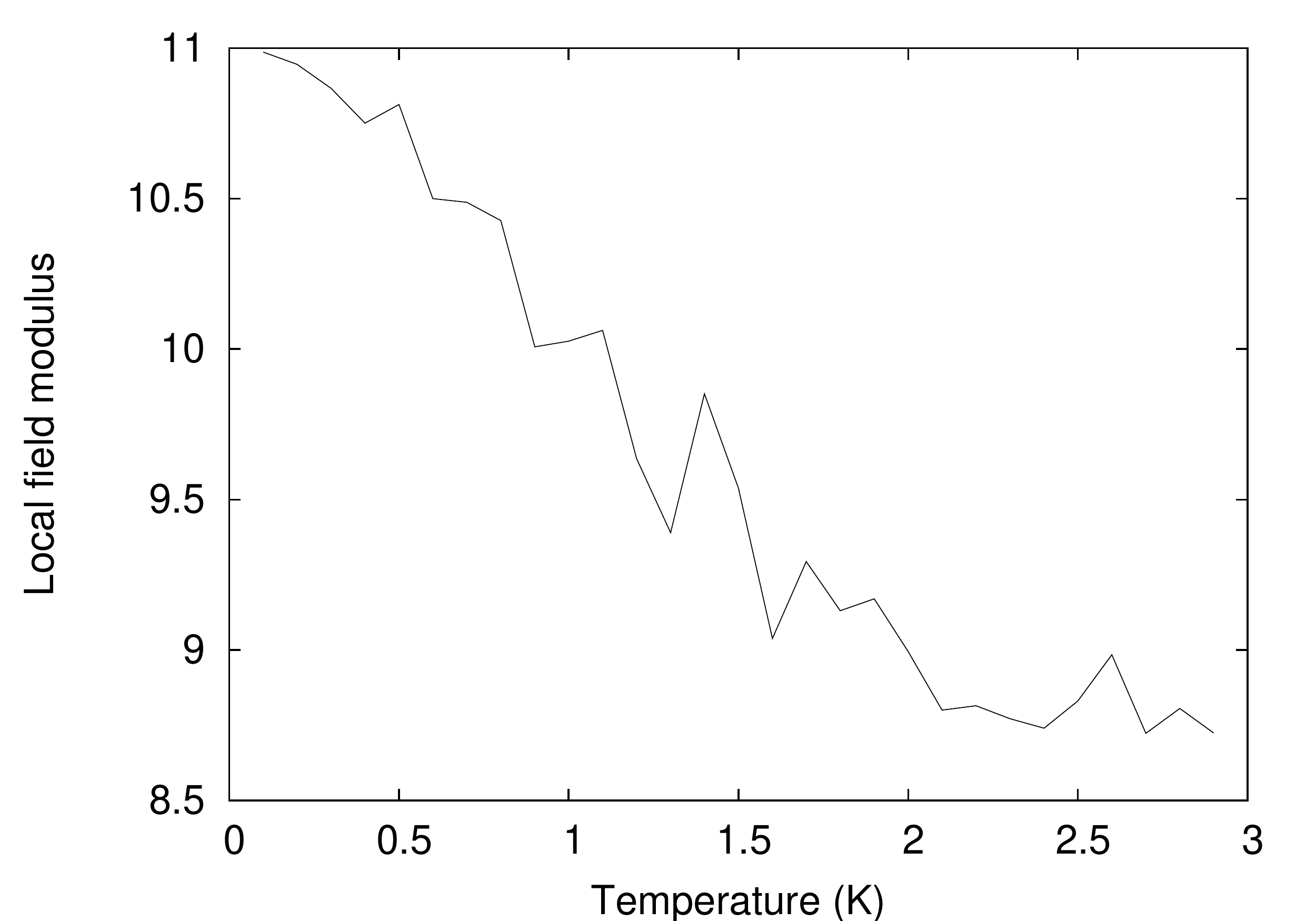}
\caption[Localfield]{{\it Modulus of Local Field as a
 function of $T$}}\label{fieldmag}
\end{center}
\end{figure}

\begin{figure}[h!]
\begin{center}
\includegraphics*[height=6.5cm]{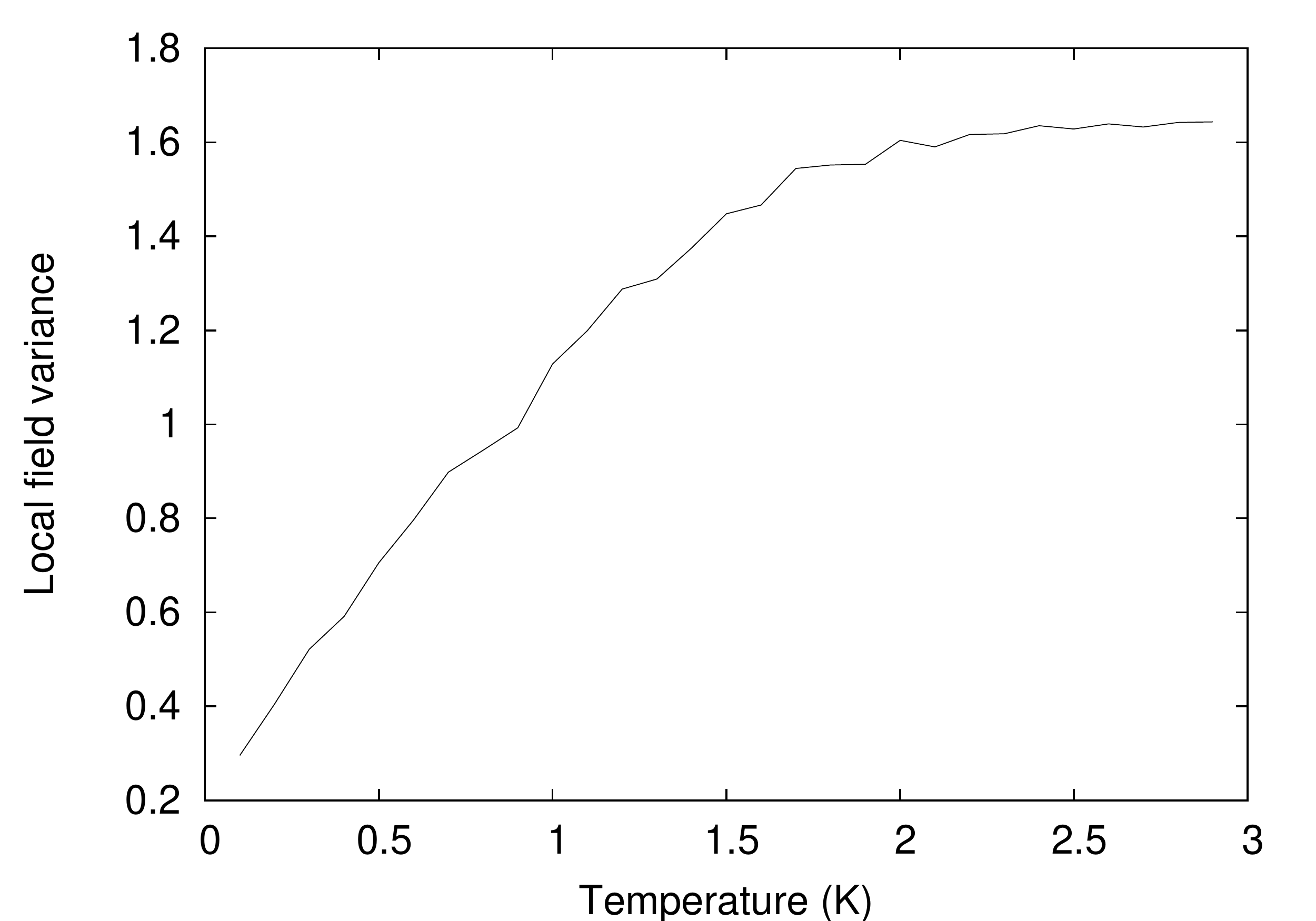}
\caption[Variance of field]{{\it Variance of Local field vs. $T$}}\label{fieldvar}
\end{center}
\end{figure}

 Fig. \ref{fieldmag} shows that the modulus of the local field on the central spin {\it increases} as $T$ lowers, while its variance, shown in Fig. \ref{fieldvar}, 
{\it decreases} rapidly for $T$ below the freezing temperature $T_f\approx 1.5-1.6\ K$ \\

\begin{figure}[h!]
\begin{center}
\includegraphics*[height=6cm]{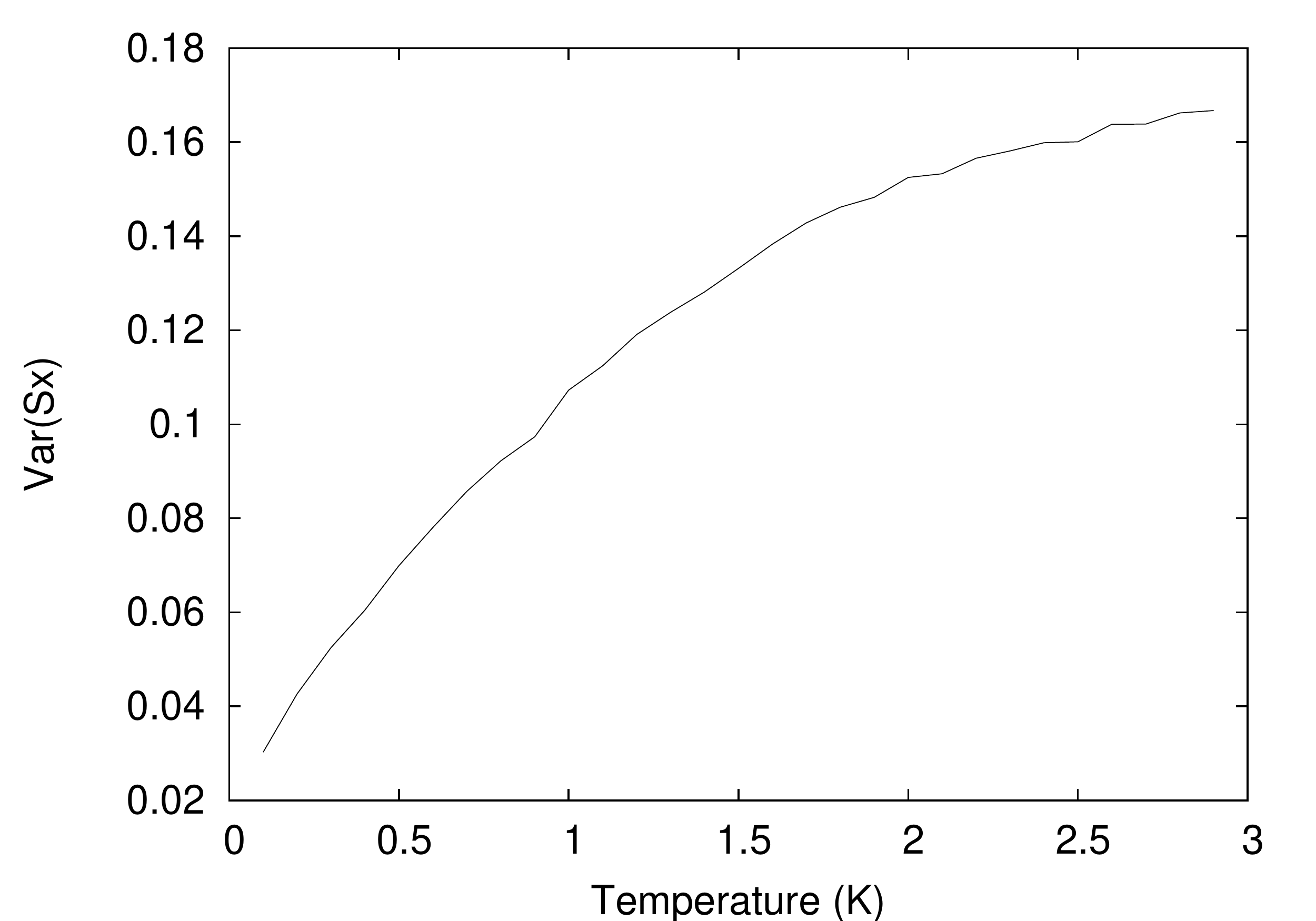}
\caption [Central Spin Variance.]{{\it Variance of $x$ component of local spin vs. $T$.}}\label{spinvar}
\end{center}
\end{figure}

The central spin exhibits a similar behaviour: Figure \ref{spinvar} shows that the variance  of the $x$ component of the central spin decreases abruptly below $T_f$. \\

We conclude that below $T_f$ each spin orientates in a given average direction, around which it fluctuates with a decreasing variance as $T$ decreases. 
This is shown by the change in the derivative shown in Fig\ref{spinvar}. On the other hand, both the specific heat (Fig.\ref{specif}) and 
the static susceptibility (Fig. \ref{sus}) show a maximum at the same $T_m$.
  
\section{Conclusions}
We present a simulation of the behaviour of a collection of nano-particles sustaining a
 magnetic dipole moment dispersed randomly in a non-magnetic film (the same conclusions are valid for
 electric dipoles, like for instance in a liquid crystal).
  We focused our study on the range of high concentrations, where
  collective behaviour can be expected.\\
  We find that such a system exhibits at low temperatures a freezing transition similar to that of a spin-glass, as shown by the temperature dependence  
of both local and global statistical properties, namely:\\
a) one can define an order parameter which is unity at very low $T$ and
decreases as $T$ increases;\\
b) each magnetic dipole at low $T$ tends to orientate in a fixed direction, around which the amplitude of its  fluctuations  decreases as $T$ lowers, which is shown by the 
correspondingly  decreasing variance and increasing average amplitude of a given spin component. Besides, the directions along which the dipoles freeze at low $T$ are random, 
which is shown by the fact that the total magnetization scales with the
number of particles as $N^{-1/2}$;\\
c) both the magnetic specific heat and the longitudinal static susceptinility show maxima at about the same temperature, which is close to the estimated freeezing temperature obtained 
by the Binder criterium based on the $T$ dependence of the kurtosis of the
statistical distribution of the magnetization;\\
d)the local magnetic field on a given spin starts increasing in amplitude and decreasing in variance as $T$ lowers below a given $T$, which coincides with the one at which the same 
properties occur for the spin components, and which we interpret as the temperature  $T_f$ of the freezing transition. \\

{\it Acknowledgements}\\

We acknowledge the invaluable help of Juan Gallego of the McGill University Physics Department with the numerical calculations and that of Adam Walters of Animetix  
for assistence with MJZ's computer system. MJZ wishes to thank {\it Compute Canada} for financial support for the use of the computer facilities of {\it Westgrid}.\\


\begin{thebibliography}{29}
\bibitem{sun}S. Sun et al., Science {\bf 287} (2000) 1989.
\bibitem{ogale}S.B.Ogale et al., Phys. Rev. Lett.  {\bf } 91 (2003) 077205.
\bibitem{allia}P.Allia et al., Phys. Rev. {\bf  52} (1995) 15398
\bibitem{mennenga}G. Mennenga et al., J. Mag. Mag. Mat. {\bf 44} (1984) 48-58.
\bibitem{altbir}D.Altbir et al., Phys. Rev. {\bf B 54} (1996) R6823.
\bibitem{edwards}F. Edwards and P. W. Anderson, J. Physique {\bf F5} (1975) 965
\bibitem{spinu}L.Spinu et al., J. Mag. Mag. Mat. {\bf 226-230} (2001) 1927
\bibitem{ohnuma}S. Ohnuma et al., IEEE Trans. on Magnetics {\bf 37} (2001) 2251.
\bibitem{held}G. A. Held, Phys. Rev. {\bf B 64} (2001) 012408
\bibitem{wagner}W. Figueiredo and W. Schwarzacher, Phys. Rev. {\bf B 77}
(2008) 104419; A.Weizenmann and W. Figueiredo, Int. J. of Mod. Phys. {\bf C
23} (2012) 124006.
\bibitem{diep}Diep The Hung et al., J. of Phys.: {\bf Conf. Series 191} (2009) 012020. 
\bibitem{Myd}J. A. Mydosh, "Spin glasses: an experimental introduction", 1993, Taylor and Francis.
\bibitem{binder} K. Binder, Zeitschrift für Physik B, Condensed Matter
June 1981, Volume {\bf 43}, Issue 2, pp. 119–140 
\bibitem{raju}J. P. Raju et al.,  Phys. Rev. {\bf B 46} (1992-I) 5405.


\end{thebibliography}
\end{document}